Surface X-ray Speckles: Coherent Surface Diffraction from Au (0 0 1)

M.S. Pierce[1], K.C. Chang[1], D. Hennessy[1], V. Komanicky[1,2], M. Sprung[3,4], A. Sandy[3], and H. You[1].

[1]Materials Science Division, Argonne National Laboratory, Argonne Illinois 60439, USA
[2]Faculty of Sciences, Safarik University, Košice, 04001, Slovakia
[3]Advanced Photon Source, Argonne National Laboratory, Argonne Illinois 60439, USA
[4]HASYLAB, DESY, Hamburg, Germany

We present coherent speckled x-ray diffraction patterns obtained from a monolayer of surface atoms. We measured both the specular anti-Bragg reflection and the off-specular hexagonal reconstruction peak for the Au (0 0 1) surface reconstruction. We observed fluctuations of the speckle patterns even when the integrated intensity appears static. By auto-correlating the speckle patterns, we were able to identify two qualitatively different surface dynamic behaviors of the hex reconstruction depending on the sample temperature.

When a system is in thermodynamic equilibrium, local properties of microstates can have considerable fluctuations away from the mean equilibrium values, even if the global average properties of the system remain unchanged. In ordinary x-ray scattering measurements, the measured x-ray intensity represents the globally averaged structure, and the information about the fluctuating microstates is completely lost and cannot be recovered. Coherent x-ray scattering[1] can in principle directly access the microstate dynamics and has been applied to study surfaces[2] and films[3]. In this Letter, we report the first application of coherent scattering to a *monolayer* of surface atoms, successfully measuring the temperature-dependent dynamics of the Au (0 0 1) surface reconstruction.

Surfaces and interfaces of metal crystals, such as Au, Pt, and Ir, present both scientifically interesting subjects and technologically important issues in heterogeneous catalysis and electrocatalysis. While nanoparticles[4,5] of transition metals, such as gold, are important in terms of real catalysts, single crystals serve as important benchmarks for our understanding of surfaces both in vacuum[6] and electrochemical conditions[7,8]. In particular, surface reconstructions and their interactions with molecules are often critical factors in determining the catalytic properties[9].

The Au (0 0 1) surface reconstructs from the face-centered-cubic arrangement to a quasi-hexagonal (hex) overlayer. The hex reconstruction of the Au (0 0 1) surface has attracted much interest since it was discovered in a Low Energy Electron Diffraction (LEED) study[6]. X-ray scattering studies[10,11] revealed a rich dependence of the surface properties upon temperature, and theoretical studies[12,13] recognized the importance of relativistic effects in atomic orbitals for the reconstruction. While the energetics of the system is well understood, very little is known about the dynamics of the hex reconstruction. In our recent studies[14], we found it is possible to measure the average kinetic behavior but impossible to directly observe the microstate dynamics using ordinary surface x-ray diffraction. The purpose of our current experimental study is twofold; i) to demonstrate a new surface coherent x-ray scattering technique and ii) to apply the technique to study the microstate dynamics of the hex phase.

These experiments were performed in reflection scattering geometry at beamline 8ID of the Advanced Photon Source at Argonne National Laboratory. Adjustable precision slits were narrowed to typically 20(h)×40(v) µm$^2$ to select photons with sufficient coherence and yet to maintain sufficient signal, illuminating an area of approximately 20×150 µm$^2$ on the sample surface. The incident x-ray flux at 7.36 keV was ~10$^{10}$ ph/sec and charge-coupled device (CCD), mounted ~2 m from the sample on the $2\theta$ arm, detected typically 10$^3$ to 10$^4$ ph/sec depending upon sample conditions. The vacuum chamber, with the Au single crystals held on a quartz pedestal in a RF induction heater, was mounted on a diffractometer, allowing sufficient rotation to access the necessary regions of reciprocal space. The samples used were single crystals cut and polished along a (001) facet, with a typical size of 5-6mm diameter. The samples, cleaned and degreased in nitric acid, were annealed first in Ar-H$_2$ for an hour and subsequently in high vacuum for several hours at ~1200 K in the vacuum chamber[14]. The temperature of the samples was measured *in situ* by determining the shift in the bulk lattice constant[15]. In order to obtain dynamic information of a thermodynamic equilibrium state, the image correlation analysis was conducted only after the integrated scattering intensity approached a constant steady-state value after the sample temperature is changed. This resulted generally in wait times of several minutes at high temperature to ~1 hour near room temperature. The rate for heating and cooling of the sample at high temperatures was usually around ~ 10 K/min.

There are two possible orientations of the hex reconstruction. These two orientations are shown in real-space in Fig. 1 (a) and the corresponding peaks in reciprocal space shown in Fig. 1 (b). CCD images collected near the (1.2 1.2 0.3) hex peak and the (0 0 1) anti-Bragg peak are shown in Fig. 2. The hex peak shown in Fig. 2(a) scatters from one of the two hex domains, while the anti-Bragg peak shown in Fig. 2(b) scatters from both hex domains. The several large spots in (a) correspond to regions of hex domains supported on several pieces of mosaic. The medium-scale faint diffraction fringes [e.g., vertical fringes on the top of (a)] represent repetitive physical domain sizes of 350(90) nm. The finest features shown in the inset is the speckle itself, a large component of which comes from the random distribution of one of the two hexagonal domains.

We were able to isolate a single mosaic domain in the case of anti-Bragg peak shown in Fig. 2(b). The speckle is evident in the fine features of the image. The scattering at the anti-Bragg condition is typically ~ 3-5 times stronger than the scattering at the hex peak, and also is an easier scattering geometry to satisfy within the limited angular ranges of the diffractometer available at the beamline. Additionally, the higher incidence angle at the anti-Bragg peak compared to the hex peak makes it easier to isolate scattering from a single mosaic block. For these reasons much of our experimental work centered on the anti-Bragg peak.

The speckle observed in the anti-Bragg peak measured at room temperature is mainly due to contributions to the phases from the height variations[16,17] due to step edges, boundaries between two hex domains, and a small (~ 7%) uniaxial corrugation within single domains due to the incommensurate nature of the surface reconstruction. As temperature increases, the hex domain boundaries move, the corrugations rearrange and, consequently, the speckles in the anti-Bragg peak evolve in a corresponding fashion. Once the hex has begun to lift at higher temperature, there is additional height variation in the (1×1) regions due to extra atoms being pushed out from the original surface layer since the topmost monolayer of the hex reconstruction has ~ 25% higher density[10,14] than that of the (1×1). Therefore, the speckle contrast would have increased when the surface reconstruction is lifted. However, the overall intensity of the anti-Bragg peak decreases by a factor of 9,[18] resulting in the speckle contrast decreased.

During the experiments, we clearly saw that at room temperature the speckle patterns were either static or very slowly changing to within our experimental uncertainty. At higher temperatures the speckles began to fluctuate and began changing at a fairly rapid rate due to the dynamic coexistence of reconstructed hex and unreconstructed (1×1) microdomains. Fig. 2(c) demonstrates the time-evolution of individual speckles in the anti-Bragg peak as a function of time for 3 different temperatures. The length of the horizontal streaks corresponds to the time over which the intensity remains relatively constant. The intensity is relatively constant at 1105 K, indicating a slow evolution, while by 1130 K the streaks nearly disappear, indicating a rapid evolution. At higher temperatures, above ~1200 K, the fluctuation of the speckles approaches the limit of our ability to measure because of the rapid rate of change, the decreased intensity, and the CCD readout time. The overall width of the anti-Bragg peak also grows from $0.6\times10^{-4}$ reciprocal lattice units (r.l.u.) to $1.0\times10^{-4}$ r.l.u. as the temperature is increased from 1105K to 1130K, indicating that the average domain size decreases.

We have applied X-ray Photon Correlation Spectroscopy (XPCS)[19,20] to extract quantitative information about the surface dynamics. In principle, real space imaging of the hex domains would be possible by performing either lensless microscopy[21,22] or holography[23]. However, for the purpose of exploring the dynamic behavior of the system, we concentrated our efforts towards XPCS. The standard time-time autocorrelation function ($g_2$) used in XPCS analysis is defined by

$g_2(t) = \frac{\langle I(t_0)I(t+t_0)\rangle}{\langle I(t_0)\rangle\langle I(t)\rangle} = 1 + \beta|f(\mathbf{Q},t)|^2$ with a symmetric normalization factor on a pixel

by pixel basis, where $\beta$ is optical contrast and $f(\mathbf{Q},t)$ is the intermediate scattering function (ISF)[24]. Selected examples of $g_2$ obtained from our data for the anti-Bragg and hex reflections are plotted in Fig. 3 (a) and (b), respectively. In order to extract quantitative rates of change, the ISF was assumed to follow a simple exponential form ($e^{-t/\tau}$). The best fits are shown by the solid black lines in Fig. 3. More complicated stretched exponentials did not improve the fits in a statistically meaningful fashion. The simple exponential decay is also justifiable within the context of dynamic scaling[25,26] for a 2-d system with a non-conserved order parameter and suggests a mean-field behavior. The time constants ($\tau$) for the exponential decay were in the range of 10 to $10^6$ sec. The contrast ($\beta$) is typically between 0.1 and 0.15 depending on the slit size for low temperatures, but decreases as the patterns begin fluctuating on rapid timescales at higher temperatures. Care was taken to ensure the values of $\tau$ used in our analysis represent equilibrium values.

The correlation time constants ($\tau$) from the results of fitting exponentials to the auto-correlation functions are plotted in Fig. 4(a). We observed that two different temperature regimes are present with ~1070 K as a transition point ($T^*$) that is shown as the vertical dashed lines in Fig. 4. At room temperature the evolution in the speckle patterns was static within the detection limit. As the temperature is increased, the speckles begin to change at a very slow rate below $T^*$, and this rate systematically increases as the temperature is raised. It is a region of slowly decreasing time constants as the temperature is increased from ambient conditions up to $T^*$. For temperatures above $T^*$, however, the speckle patterns begin decorrelating on much faster timescales. We observe the time constant to rapidly drop as the temperature is further increased. The integrated intensities of the hex peak, shown in Fig. 4(b), also began to drop at $T^*$. The values of $\tau$ measured at the anti-Bragg and at the hex peak both are shown in Fig. 4(a). The values measured at the hex peak are consistent with the values obtained from anti-Bragg data, though they are systematically smaller. We believe the smaller time constants can be attributed to the hex peak's sensitivity to lateral motion of hex domain boundaries.

We assume that the inverse of the correlation time $\tau$, the rate of change, should follow the simple Arrhenius relation, $\frac{1}{\tau} \propto e^{-E/kT}$, where $E$ is the activation energy. It is shown[11] that hex domains grow slowly over 24 hours. This indicates that the boundaries between the two hex domains are slowly moving and changing with time at temperature between 435 and 1200 K. Since the energy of the two hex domains are the same, the boundary motion is governed by the activation barrier ($E_a$), and $\frac{1}{\tau} \propto e^{-E_a/kT}$. However, at high temperatures close to $T_c$, we expect that activation of (1×1) domains dominates the fluctuation of the microstates. The (1×1) activation is governed by the free energy difference ($\Delta G$) between hex and (1×1). While we can assume that $E_a$ is insensitive to temperature, $\Delta G$ tends rapidly to zero as $T$ approaches $T_c$. In the absence of detailed analytic models for the hex reconstruction, we invoke a Landau-form free energy for the hex to (1×1) transition, $G(\psi_{hex}) = (T - T_c)\frac{a}{2}\psi_{hex}^2 + \frac{b}{4}\psi_{hex}^4 + \cdots$, to obtain $\Delta G \approx \alpha k (T - T_c)^2 / T_c$ where $\alpha = \frac{3a^2 T_c}{4b}$. This result states simply that the free energy difference becomes smaller as $T$ approaches $T_c$ and the transition between them becomes increasingly easier. Then, the expression for $\tau$ from the sum of these two effects can be written as,

$$\frac{1}{\tau} = v_1 e^{-E_a/kT} + v_2 e^{-\alpha(T_c-T)^2/T_c T}$$  Eq. (1)

where $v_1$ and $v_2$ are proportional constants. The best fit using this equation is shown as the solid line in Fig. 4(c) with the fit parameters $E_a = 0.15(2)$ eV and $\alpha = 860(80)$. We can see that the first term dominates below $T^*$ and the second term dominates above $T^*$. Below $T^*$, therefore, we believe the surface phases are transitioning mainly between the two different populations of rotated hex phases. Above $T^*$, the surface reconstruction begins to lift, and the dynamics are dominated by the transition of hex domains to the (1×1) surface.

In principle, we must consider one additional term in Eq. (1) if we can measure the shorter time scales since the motion of (1×1) islands is expected to be fast and dominate the dynamics around and above $T_c$. However, we reached our time-resolution limit near $T_c$ and were unable to obtain accurate time constants. Nevertheless, the time constants we attempted to measure above $T_c$ are shown in Fig. 4 (a) and (c) as filled squares. They have a rather large error bars and several times smaller than the reliable time constants measured below $T_c$. We expect that the time resolution will improve as both detector efficiency and x-ray brilliance increase in the future.

We also collected speckle patterns of the bulk reflection (002). At high temperatures, there does appear to be fluctuation in the speckles, but it occurs at a much slower rate. At temperatures of 1100 K, we find the time constants extracted from the bulk (002) reflection to be ~ $10^5$ sec., a rate of ~ 100 times slower than observed at the anti-Bragg condition. This is further evidence that the observed phenomena at both the anti-Bragg peak and the hex peak are in fact surface dynamics and not due to changes within the crystal bulk.

In conclusion, we have demonstrated coherent surface x-ray scattering from a single surface monolayer of Au (0 0 1) and found that the rates of the microstate evolution exhibit the two qualitatively different dynamic behaviors below $T_c$. At low temperatures, we observe slow evolutions of the surface mainly due to the motion of hex domain boundaries. Above the temperature at which the hex begins to lift, we observe acceleration in dynamic behavior due to the narrowing of the free energy difference between the hex and (1×1) phases. We observe these changes in real time, even though the average macrostate of the system has stopped evolving. This provides new dynamic information that we hope in the future can be applied to further our understanding of surface phase transitions and gas-surface interaction dynamics.

This work and use of the Advanced Photon Source were supported by the U. S. Department of Energy, Office of Science, Office of Basic Energy Sciences, under Contract No. DE-AC02-06CH11357.

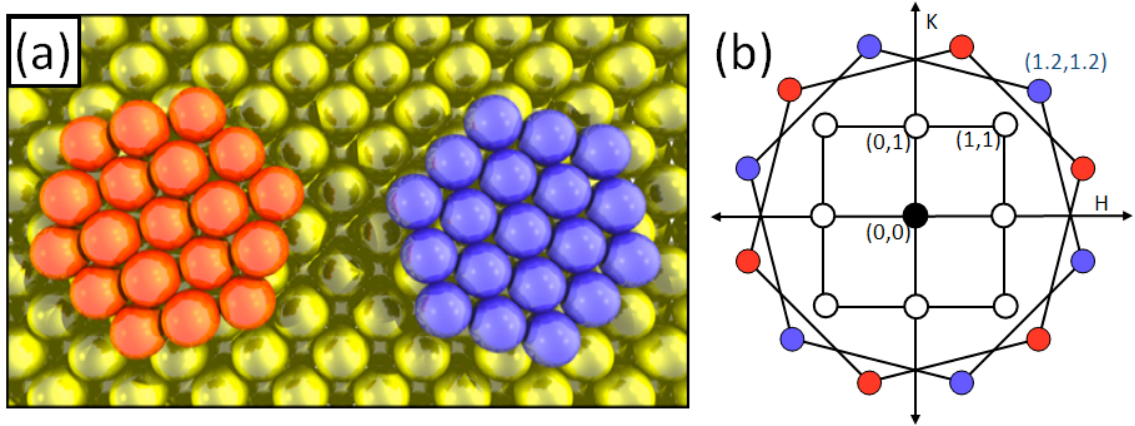

Figure 1. (a) A representation of the two possible real-space orientations for the hexagonal lattice over the (0 0 1) bulk. (b) The reciprocal space diagram for the two orientations at L=0.

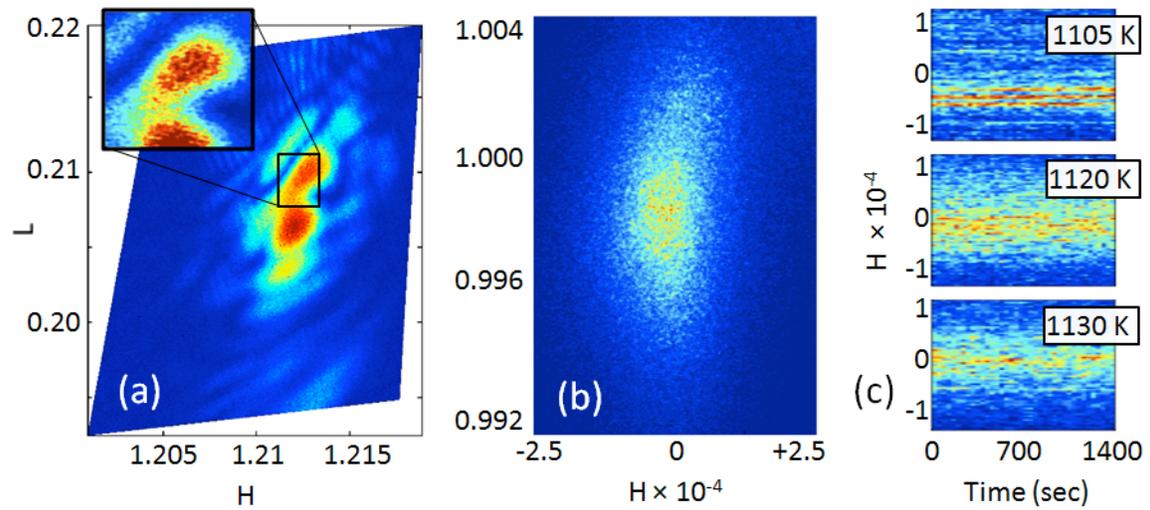

Figure 2. Scattering patterns collected at room temperature for the hex peak (a) and the anti-Bragg peak (b). Overlap of a few mosaic blocks was unavoidable at glancing angles in (a); a single mosaic block was isolated in (b). (c) Single pixel slices of the speckle patterns from the anti-Bragg peak plotted as a function of time at 1105K, 1120K, and 1130K, respectively. Reciprocal lattice units are used for H and L.

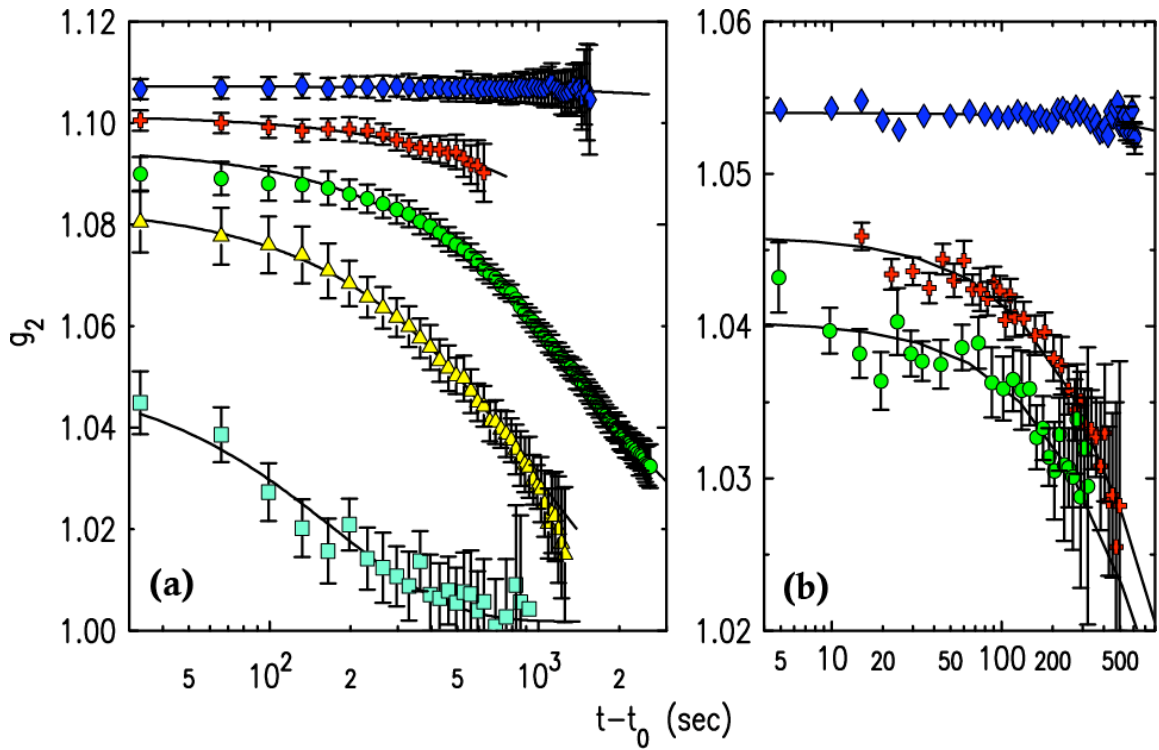

Figure 3. (color online) Selected auto-correlation measured at (a) the anti-Bragg condition and (b) hex reflection. The solid lines indicate exponential fits. The temperatures are 300 K (◊), 1080 K (✣), 1105 K (○), 1120 K (Δ), 1140 K (□), respectively. The top two auto-correlations (◊ and ✣) in (a) are both offset for clarity by 0.015 and 0.01, respectively.

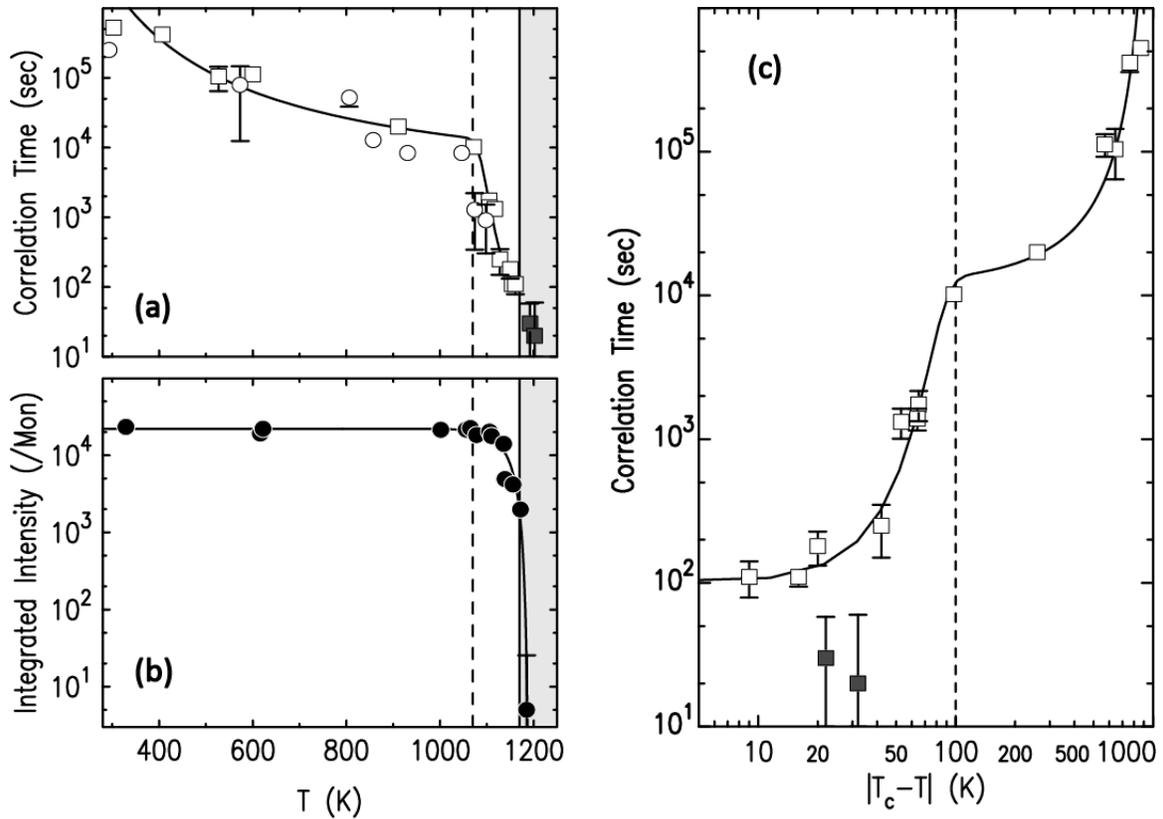

Figure 4 Time constants (a), $\tau$, extracted from auto-correlation fits to both the hex (open circles) and anti-Bragg peaks (squares), and integrated hex peak intensity (b) are plotted vs. $T$. The solid lines are guides to eye. The $T > T_c$ region is shaded for clarity. (c) The time constants from the anti-Bragg peak are re-plotted vs. $|T_c-T|$. The solid line is the fit using Eq. (1). The vertical dash lines indicate $T^*$. The two filled squares were measured above $T_c$.